\documentclass[aps, pra, reprint, twocolumn, superscriptaddress, floatfix]{revtex4-2}

\usepackage[T1]{fontenc}
\usepackage[utf8]{inputenc}
\usepackage{amsmath,amssymb,bm}
\usepackage{graphicx}
\usepackage{booktabs}
\usepackage{multirow}
\usepackage{array}
\usepackage{xcolor}
\usepackage{placeins}
\usepackage[colorlinks=true,linkcolor=blue,citecolor=blue,urlcolor=blue]{hyperref}
\usepackage{bibunits}

\newcommand{\dd}{\mathrm{d}}
\newcommand{\Tr}{\operatorname{Tr}}
\newcommand{\avg}[1]{\langle #1 \rangle}
\newcommand{\ket}[1]{|#1\rangle}
\newcommand{\bra}[1]{\langle #1|}
\newcommand{\braket}[2]{\langle #1 | #2 \rangle}
\newcommand{\Vnc}{V_{\rm nc}}
\newcommand{\nbar}{\bar{n}}
\newcommand{\result}[2]{\par\smallskip\noindent\parbox{\linewidth}{\textbf{#1.}\ \textit{#2}}\par\smallskip}

\graphicspath{{figures/}}

\begin{document}

\title{Complementary quantum and classical records of qubit decoherence}

\author{Jargalsaikhan Artag}
\author{Koki Awaya}
\author{Takumi Kanezashi}
\author{Haruya Nagata}
\author{Daisuke Tsukayama}
\author{Moe Shimada}
\author{Jun-ichi Shirakashi}
\affiliation{Department of Electrical Engineering and Computer Science, Tokyo University of Agriculture and Technology, Koganei, Tokyo, Japan}

\date{\today}

\begin{abstract}
Decoherence is usually viewed as the loss of local coherence, but it also writes information into the environment. Here we examine two complementary forms of this information. One is a recoverable quantum record: conditioning on a transverse qubit measurement reveals a Schr\"odinger-cat-like state in an occupation-selected collective bath mode. The other is a redundant classical which-path record distributed across physical frequency-band fragments. Using tensor-network simulations of the spin--boson model, we reconstruct the conditional bath Wigner function and resolve its negativity. At zero temperature, one natural orbital carries more than $97\%$ of the bath one-body occupation in the chain representation. For an initial $\ket{+x}$ qubit and a pure parity-even bath, a symmetry gives an exact nonperturbative identity between the remaining qubit coherence and the overlap of the two environmental branches. Exact pure-dephasing calculations establish both conditional nonclassicality and redundant pointer information in the same system--bath state. With tunneling, a detector that excludes frequencies below $0.1\omega_c$ resolves transient conditional negativity at the simulated temperatures $T/\omega_0=0.3,0.5,0.7,1$; its suppression reflects both uncollected modes and thermal effects. In pure dephasing, the conditional state of the displacement-matched mode is an exact thermal mixture of displaced cat states. These results connect decoherence to phase-space tomography and motivate a readout combining qubit measurement and collective-mode Wigner tomography.
\end{abstract}

\maketitle

\begin{bibunit}[apsrev4-2]

\section{Introduction}
Decoherence is often summarized as a loss: a local system loses phase coherence when it becomes entangled with its environment~\cite{Zurek2003,Schlosshauer2007,Weiss2012,Breuer2002}. That summary is incomplete. Under unitary system--environment evolution, the missing local coherence is encoded in their joint state. The physical question is therefore not only how fast a qubit decoheres, but what records the environment receives while it does so.

The usual theoretical tools---reduced dynamics, influence functionals, and master equations~\cite{Feynman1963,CaldeiraLeggett1983,Breuer2002}---remove the environment to predict relaxation and dephasing. Quantum Darwinism instead asks how separate fragments of an environment acquire repeated copies of classical pointer information~\cite{Zurek2003,OllivierPoulinZurek2004,Zurek2009Darwinism}. This classical record is only part of the story. Coherence between the environmental branches can still be recovered by measuring the qubit in a transverse basis and selecting the environmental state associated with one outcome. We refer to this selection as \emph{conditioning}~\cite{Carmichael1993,WisemanMilburn2009,Murch2013}. It exposes a second, quantum record: the branch superposition hidden when the qubit is simply traced out.

Here we study two complementary records produced by qubit decoherence in the same environment. The quantum record is recovered as a nonclassical state in an occupation-selected collective bath mode. The classical which-path record is distributed across physical frequency-band fragments. We demonstrate both records in the same system--bath state in pure dephasing and reconstruct the conditional nonclassical state with finite tunneling. This separates the operational questions of recovering a coherent branch superposition and reading repeated copies of pointer information.

\begin{figure}[tb]
\centering
\includegraphics[width=\columnwidth]{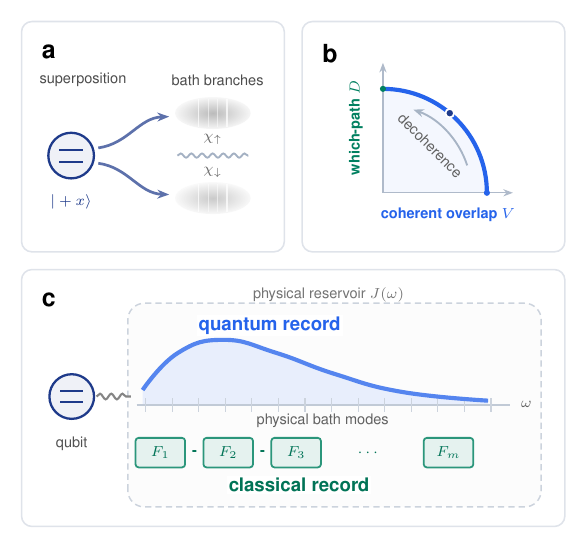}
\caption{\textbf{Complementary records emerge in the environment.} \textbf{(a)} A qubit superposition evolves into two correlated bath branches; tracing out the bath is what appears locally as coherence loss. \textbf{(b)} The qubit coherence equals the normalized environmental branch overlap, while the which-path distinguishability $D$ grows as the coherent overlap $V$ falls. At zero temperature their squared values sum to one through the exact identity and the visibility--distinguishability relation stated as Theorems~1 and 2. \textbf{(c)} The quantum record is read, after a transverse qubit measurement, from one time-dependent collective bath mode, a coherent superposition of the physical frequency modes. The classical which-path record is read from physical frequency-band fragments $F_1,F_2,\ldots,F_m$ of the same modes; the two records are accessed by different measurements, not stored in separate places.}
\label{fig:concept}
\end{figure}

Structured, non-Markovian reservoirs provide a demanding setting in which to test this dual-record picture~\cite{Breuer2009,Rivas2014,deVega2017,Li2018}. They have memory, strong frequency dependence, and many environmental degrees of freedom. We study the canonical spin--boson model~\cite{Leggett1987,Weiss2012}, using a chain mapping and tensor-network evolution to reach strong coupling and finite temperature~\cite{Chin2010,Woods2014,Prior2010,Tamascelli2018,Tamascelli2019}. A transverse qubit measurement reveals a Wigner-negative state in an occupation-selected collective bath mode. Exact pure-dephasing calculations separately quantify repeated which-path information in physical frequency-band fragments. For an initial $\ket{+x}$ qubit and a pure parity-even bath, a parity symmetry equates the magnitude of the environmental branch overlap with the qubit coherence. Wigner negativity tests the separate question of whether the selected environmental state is nonclassical~\cite{Wigner1932,Hudson1974,Kenfack2004}. The proposed readout builds on quantum-optical and circuit-QED tomography~\cite{Vogel1989,Banaszek1996,Lutterbach1997,Lvovsky2009,Deleglise2008,Vlastakis2013}, with capture of the collective mode as an additional requirement. Finally, we examine how thermal mixing changes this quantum record, analytically in pure dephasing and numerically with tunneling. Figure~\ref{fig:concept} summarizes the two records and the measurements that distinguish them.

\section{A cat in a collective bath mode}
We reconstruct the recoverable quantum record as a Schr\"odinger-cat-like state in one occupation-selected, time-dependent collective coordinate. To study this state in a nonperturbative reservoir, we take the unbiased spin--boson Hamiltonian
\begin{equation}
H_\star=\Delta\sigma_x+\sum_k\Omega_k b_k^\dagger b_k
+\sigma_z\sum_k\lambda_k(b_k^\dagger+b_k),
\label{eq:star}
\end{equation}
with tunneling amplitude $\Delta$, bath modes $b_k$ of frequency $\Omega_k$, and spectral density
\begin{equation}
J(\omega)=2\alpha\,\omega_c^{1-s}\omega^s e^{-\omega/\omega_c},\qquad\omega>0,
\label{eq:Jw}
\end{equation}
of coupling $\alpha$, cutoff $\omega_c$, and exponent $s$. With $J(\omega)=\sum_k\lambda_k^2\delta(\omega-\Omega_k)$, Eq.~\eqref{eq:Jw} fixes our convention for $\alpha$. In the standard numerical-renormalization-group (NRG) normalization, it corresponds to $\alpha_{\rm std}=4\alpha$ (Methods), so the primary simulated coupling $\alpha=0.4$ is $\alpha_{\rm std}=1.6$, a strongly coupled, nonperturbative regime. The isolated qubit transition has energy $2\Delta$. We use a fixed frequency unit $\omega_0=\omega_c/4$ and $\hbar=k_B=1$, reporting time as $t\omega_0$ and temperature as $T/\omega_0$. The tunneling simulations use $\Delta=\omega_0$, $s=0.5$, and $\alpha=0.4$ unless stated otherwise, and probe $T/\omega_0=0,0.3,0.5,0.7,1$. This unit remains defined in the pure-dephasing limit $\Delta=0$. The initial state is $\ket{+x}\bra{+x}\otimes\rho_B^{\rm th}$.

\begin{figure*}[!t]
\centering
\includegraphics[width=0.95\textwidth]{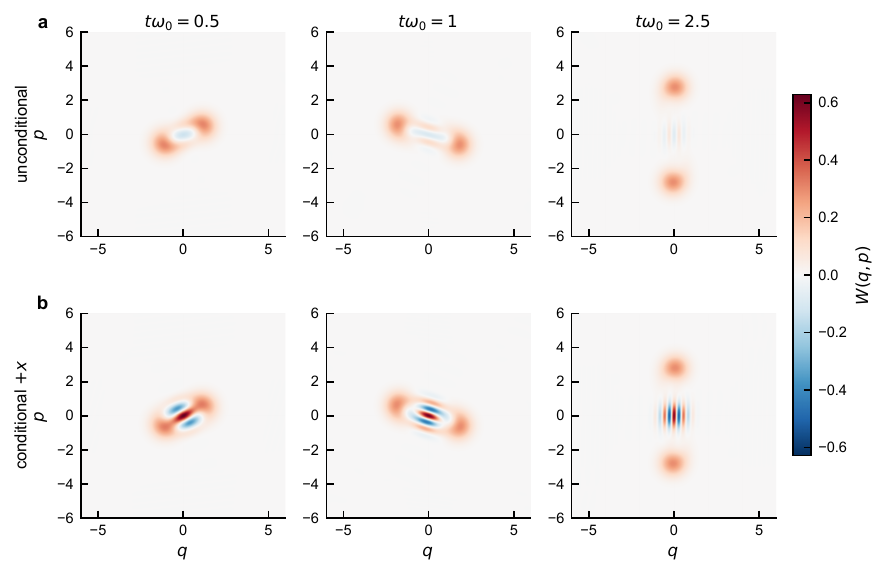}
\caption{\textbf{A Schr\"odinger-cat-like state emerges in a collective bath mode.} Wigner function of the occupation-selected collective bath mode, without a detector band, for $s=0.5$, $T=0$, $\alpha=0.4$, $\Delta=\omega_0$. \textbf{(a)} Unconditional mode: the qubit result is ignored, producing a mixture of the two branches, with weaker nonclassicality. \textbf{(b)} Measurement-conditioned mode: selecting the $+x$ qubit result reveals two lobes with negative interference fringes (blue), the signature of a Schr\"odinger-cat-like state.}
\label{fig:emergence}
\end{figure*}

The bath is mapped by an exact change of basis to a chain of bosonic oscillators~\cite{Chin2010,Woods2014},
\begin{equation}
\begin{aligned}
H={}&\Delta\sigma_x+\sum_{n}\omega_n c_n^\dagger c_n
+\sum_{n}t_n(c_n^\dagger c_{n+1}+\mathrm{h.c.})\\
&+g\,\sigma_z(c_0^\dagger+c_0),
\end{aligned}
\label{eq:chain}
\end{equation}
where $g$ is the coupling strength between the qubit and the first chain mode $c_0$. We evolve $\ket{\Psi(t)}$ with the two-site time-dependent variational principle (TDVP) and reconstruct, at each time, the leading natural orbital $c_f(t)=\sum_k f_k(t)c_k$ of the bath one-body density matrix $C_{mn}=\avg{c_m^\dagger c_n}$. At zero temperature this chain-basis orbital maximizes the physical one-body occupation among normalized linear modes (a time-dependent natural orbital, related to but distinct from the fixed reaction coordinate of chain mappings~\cite{Martinazzo2011,IlesSmith2014,Nazir2018}). Writing the eigenvalues of this density matrix as $\lambda_j$, one orbital carries almost all of the cold one-body occupation, $\lambda_1/\sum_j\lambda_j>0.97$ at the sampled times (Fig.~\ref{fig:Smode}); in cold pure dephasing the ratio is exactly $1$ once the bath is occupied. This quantifies occupation concentration; Wigner negativity independently establishes nonclassicality in the selected mode. Figures~\ref{fig:emergence} and~\ref{fig:law} use this orbital, selected from the unconditional state and used unchanged after conditioning. For the temperature comparison in Fig.~\ref{fig:thermal}, the occupation-derived profile is passed through a physical detector band defined in Sec.~\ref{sec:thermal}. At finite temperature, selection in the thermalized chain does not maximize the occupation of the physical bath (Supplementary Sec.~\ref{si:mode}).

The object of interest is the bath state selected by a transverse qubit measurement. We call it the \emph{conditional} state because it is conditioned on obtaining the $+x$ outcome. For a pure global state and $\Pi_{+x}=\ket{+x}\!\bra{+x}$,
\begin{equation}
\ket{\psi_B^{(+x)}(t)}=\frac{\braket{+x}{\Psi(t)}}{\sqrt{P_{+x}(t)}},\qquad
P_{+x}(t)=\avg{\Pi_{+x}}_\Psi ,
\label{eq:cond}
\end{equation}
defines this normalized state and its measurement probability $P_{+x}$. By contrast, the \emph{unconditional} state $\rho_B=\Tr_S\ket{\Psi}\!\bra{\Psi}$ ignores the qubit outcome and therefore mixes the two environmental branches. An observer with access only to the environment sees this incoherent mixture: the quantum record requires the transverse qubit outcome, communicated classically, whereas the classical which-path record can be read from bath fragments alone. For a pure global state, written in the pointer ($\sigma_z$) basis as
\begin{equation}
\ket{\Psi(t)}=c_\uparrow(t)\ket{\uparrow}\ket{\chi_\uparrow(t)}
+c_\downarrow(t)\ket{\downarrow}\ket{\chi_\downarrow(t)},
\label{eq:branch}
\end{equation}
the $+x$ projection prepares the branch superposition $\propto c_\uparrow\ket{\chi_\uparrow}+c_\downarrow\ket{\chi_\downarrow}$---a Schr\"odinger-cat-like superposition of the two environmental branches. At finite temperature the physical conditional state is instead
\begin{equation}
\rho_B^{(+x)}=\frac{\bra{+x}\rho_{SB}\ket{+x}}{P_{+x}},\qquad
P_{+x}=\Tr_B\bra{+x}\rho_{SB}\ket{+x}.
\label{eq:mixedcond}
\end{equation}

Figure~\ref{fig:emergence} shows that tracing out the qubit hides the quantum record, while conditioning reveals it in phase space. For $s=0.5$, $T=0$, $\alpha=0.4$, the unconditional mode [Fig.~\ref{fig:emergence}(a)] is a mixture of the two branches with weaker nonclassicality. Conditioning [Fig.~\ref{fig:emergence}(b)] reveals two separating lobes with clear interference fringes---a cat-like state with Wigner negativity. Writing its Wigner function as $W(q,p;t)$, we quantify the negative phase-space weight by the negativity volume~\cite{Kenfack2004}
\begin{equation}
\Vnc(t)=2\iint_{W<0}\dd q\,\dd p\,|W(q,p;t)|.
\label{eq:Vnc}
\end{equation}

We write $\Vnc^{(+x)}$ for the conditional state. We use dimensionless phase-space coordinates $z=q+ip$; the vacuum Wigner function is $W_0(q,p)=2e^{-2(q^2+p^2)}/\pi$, normalized to $\iint W_0\,\dd q\,\dd p=1$.

Supplementary Video~1 shows the conditional Wigner evolution of this cold mode; Videos~2 and~3 show the detector band of Fig.~\ref{fig:thermal} at $T/\omega_0=0.5$ and $1$.

\section{Complementary coherent and which-path records}
The two environmental records require distinct diagnostics. The branch overlap controls the residual qubit coherence and the distinguishability of the pure bath branches. Conditional Wigner negativity tests the nonclassicality of a selected bath mode, while fragment information quantifies repeated pointer records. From Eq.~\eqref{eq:branch},
\begin{equation}
\avg{\sigma_x}(t)=2\,\mathrm{Re}\!\left[\bar c_\uparrow(t)c_\downarrow(t)\,
\braket{\chi_\uparrow(t)}{\chi_\downarrow(t)}\right],
\label{eq:law}
\end{equation}
the qubit coherence is set by the overlap of the two environmental branches and their populations. The unbiased model has a conserved parity symmetry $\Pi=\sigma_x\otimes P$, where $P$ reverses every bath coordinate. Starting from the parity-even state used here, this symmetry keeps the two branch populations equal. It also makes their phases cancel in Eq.~\eqref{eq:law}. This turns the record picture into an exact statement: the local coherence remaining in the qubit is precisely the coherent overlap still present between the environmental branches. Supplementary Sec.~\ref{si:proofs} gives the proof.

\result{Exact identity (Theorem 1: coherence--overlap)}{For the unbiased spin--boson model evolving from $\ket{+x}$ and a pure parity-even bath state, the qubit coherence equals the overlap of the two environmental branches at all times, couplings, and tunneling amplitudes $\Delta$: $|\avg{\sigma_x}(t)|=|\braket{\chi_\uparrow(t)}{\chi_\downarrow(t)}|$.}

\noindent At finite temperature, the branch-ket identity refers to a parity-compatible purification. In the physical mixed bath, define the unnormalized off-diagonal block $X_B=\bra{\uparrow}\rho_{SB}\ket{\downarrow}$. Parity invariance gives $V=2|\Tr_B X_B|=|\avg{\sigma_x}|$, which is not the fidelity of the physical mixed branches.

For pure bath branches, the same overlap also determines how well the environment distinguishes the two qubit pointer states. We denote the qubit visibility by $V=|\avg{\sigma_x}|$ and the which-path distinguishability by $D$, the trace distance between the two bath branch states. A large $V$ means that the branches still overlap. A large $D$ means that a measurement of the full bath can distinguish them~\cite{WoottersZurek1979,Jaeger1995,Englert1996}.

\result{Visibility--distinguishability relation (Theorem 2)}{At zero temperature the coherent branch overlap $V=|\avg{\sigma_x}|$ and the environment's which-path distinguishability $D$ obey $V^2+D^2=1$.}

\noindent This is the pure-state visibility--distinguishability relation~\cite{WoottersZurek1979,Jaeger1995,Englert1996}, applied here along the reservoir evolution. Decoherence thus converts coherent overlap into which-path distinguishability. As $V$ decreases, $D$ increases, and their squared values always sum to one at zero temperature. This relation concerns the complete pure bath branches; it does not identify Wigner negativity with visibility or specify where information is stored.

The nonclassical record is observed in a single collective mode. In the cold chain representation, one collective orbital carries about $97\%$ of the bath occupation and contains the recoverable cat-like state. In pure dephasing, the classical record is \emph{redundant}. To quantify it, we express the bath in physical frequency modes and ask how much information a frequency-band fragment contains about the qubit pointer state. We use the Holevo information, which bounds accessible classical information, for this test. Full definitions and formulas are given in Supplementary Sec.~\ref{si:darwin}.

We define the redundancy $R_\delta=M/m_\delta$, where $m_\delta$ is the smallest number of the $M=128$ physical frequency bands whose average Holevo information reaches a fraction $1-\delta$ of the whole-bath value, with $\delta=0.1$; this average-fragment estimate does not certify a partition into $R_\delta$ disjoint successful fragments. In pure dephasing at $T=0$ and $t\omega_0=2.5$, random fragments give $R_\delta\approx12.8$, while conditioning the same system--bath state yields an unfiltered collective mode with $\Vnc\approx0.631$: the redundant classical record and the conditional cat are present at the same time. Supplementary Sec.~\ref{si:darwin} and Figs.~\ref{fig:Sdarwin} and~\ref{fig:Sdarwinrobust} give the definitions, the thermal information bounds, and the partition dependence. With tunneling, the physical branch states need not be displaced Gaussians, so fragment information is quoted only at $\Delta=0$. The cold tunneling calculation still leaves a residual coherence near $0.14$ and a visible conditional cat.

\begin{figure}[!t]
\centering
\includegraphics[width=\columnwidth]{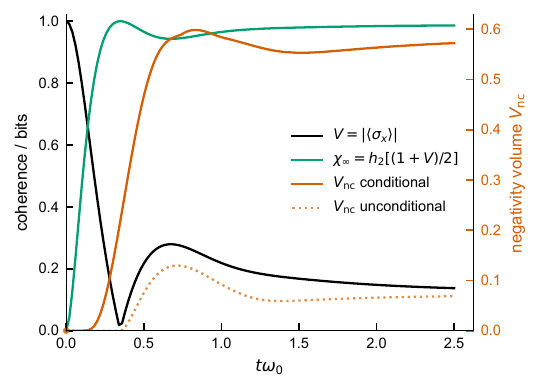}
\caption{\textbf{Qubit coherence equals the bath-branch overlap.} Black: qubit coherence $V=|\avg{\sigma_x}|$. The independently evaluated branch overlap $|\braket{\chi_\uparrow}{\chi_\downarrow}|$ coincides with it (Theorem~1; residual in Supplementary Sec.~\ref{si:wigner}). Green: whole-bath Holevo information $\chi_\infty=h_2[(1+V)/2]$ of the pure cold branches. Right axis: conditional vs.\ unconditional negativity volume $\Vnc$ of the unfiltered occupation-selected mode, showing that conditioning enhances the nonclassicality of the selected mode. Here $\Delta=\omega_0$, $T=0$, $s=0.5$, and $\alpha=0.4$.}
\label{fig:law}
\end{figure}

In Fig.~\ref{fig:law}, the independently contracted branch overlap coincides with $V=|\avg{\sigma_x}|$ to $2.4\times10^{-15}$ (Supplementary Sec.~\ref{si:wigner}). For the pure cold branches, the whole-bath Holevo information of the pointer states is $\chi_\infty=h_2[(1+V)/2]$, where $h_2(x)=-x\log_2x-(1-x)\log_2(1-x)$ is the binary entropy. Its rise to nearly one bit quantifies the total pointer information available in the bath, without determining its distribution among fragments. The same plot contrasts the conditional and unconditional negativity volumes: $\Vnc^{(+x)}$ is large (peak $\approx0.60$) while the unconditional $\Vnc$ is several-fold smaller (peak $\approx0.13$), showing that conditioning makes additional nonclassicality accessible. The outcome probability $P_{+x}=(1+\avg{\sigma_x})/2$ stays between about $0.36$ and $0.49$ after the first coherence minimum, so the conditional state is prepared in a substantial fraction of runs.

\section{A persistent record at zero temperature}

\begin{figure*}[!t]
\centering
\includegraphics[width=\textwidth]{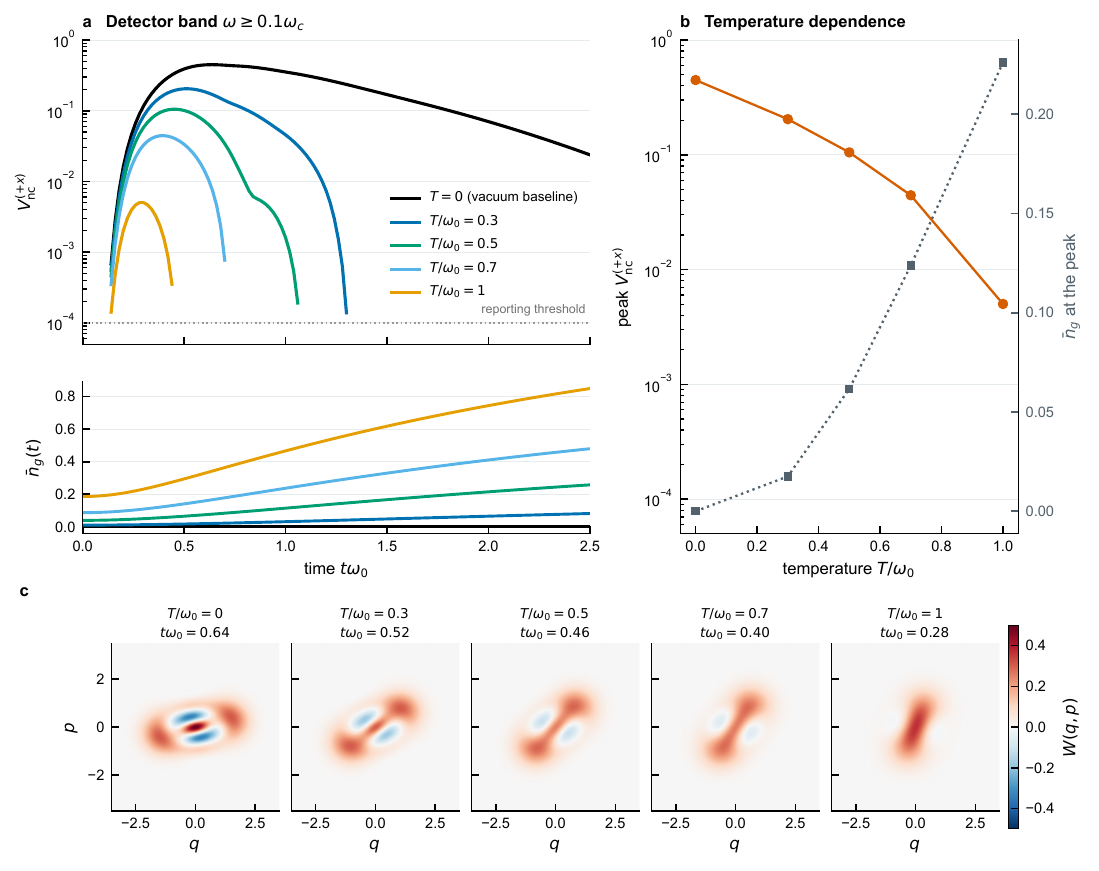}
\caption{\textbf{Finite-temperature conditional negativity through a common detector band.} All curves use $\Delta=\omega_0$, $s=0.5$, $\alpha=0.4$, and the occupation selector filtered to $\omega\ge0.1\omega_c$ [Eq.~\eqref{eq:detector}]. \textbf{(a)} Conditional negative volume versus time, on a logarithmic scale, for the vacuum baseline $T=0$ and for $T/\omega_0=0.3$, $0.5$, $0.7$, and $1$; values below the $10^{-4}$ reporting threshold are not shown. The lower panel gives the thermal background $\nbar_g(t)$ of the detected mode. \textbf{(b)} Peak conditional negativity versus temperature, with $\nbar_g$ at the same time. \textbf{(c)} Conditional Wigner function of the detected mode at the time of peak negativity for each temperature, on a shared linear color scale. The unfiltered cold mode of Fig.~\ref{fig:law} keeps $\Vnc\approx0.57$ at $t\omega_0=2.5$.}
\label{fig:thermal}
\end{figure*}

At zero temperature the negative Wigner volume persists over the simulated interval and is observed across the couplings considered. For $s=0.5$, $\alpha=0.4$ the conditional record reaches $\Vnc^{(+x)}\approx0.60$, with a late-time minimum Wigner value $W_{\min}\approx-0.51$, and remains near its peak after formation (Fig.~\ref{fig:law}). In the pure-dephasing ($\Delta=0$) limit the matched-mode result is analytic; with tunneling the stated persistence is limited to the simulated window. At $t\omega_0=2.5$, the unfiltered cold conditional mode has fitted even-cat fidelity $0.930$ and purity $0.870$; these quantify a cat-like mixed mode rather than an ideal pure cat (Supplementary Sec.~\ref{si:wigner}). This persistence refers to the unfiltered collective mode. For the same bath trajectory, a detector that collects only $\omega\ge0.1\omega_c$ gives a peak $\Vnc=0.448$ near $t\omega_0=0.64$ and a value of $0.024$ at $t\omega_0=2.5$ (Fig.~\ref{fig:thermal}). The reduced state of this filtered mode loses most of its negative volume when the low-frequency band is left uncollected. In the pure-dephasing solution, these low-frequency modes gain branch displacement over this interval, and tracing them out suppresses the detected-mode interference [Supplementary Eq.~\eqref{eq:Sfragmentcontrast}].

This behavior is not confined to one coupling. The peak negativity rises only mildly with $\alpha$, from $\Vnc^{(+x)}\approx0.53$ at $\alpha=0.2$ to $\approx0.61$ at $\alpha=0.5$ (Fig.~\ref{fig:Salpha}). The effect is therefore present across the tested couplings rather than at a finely tuned point. Stronger coupling separates the two lobes faster and produces a modestly larger fringe volume.

The chain-basis profile $f_k(t)$ of the leading orbital moves outward along the mapped chain (Fig.~\ref{fig:Smode}). The Wigner reconstruction independently shows that the selected, time-dependent mode remains nonclassical over the simulated interval. In pure dephasing this mode follows the exact cat solution. The fragment analysis in Supplementary Fig.~\ref{fig:Sdarwin} establishes the redundant classical record in the same pure-dephasing bath.

\section{The record at finite temperature}
\label{sec:thermal}

\begin{figure*}[!t]
\centering
\includegraphics[width=0.95\textwidth]{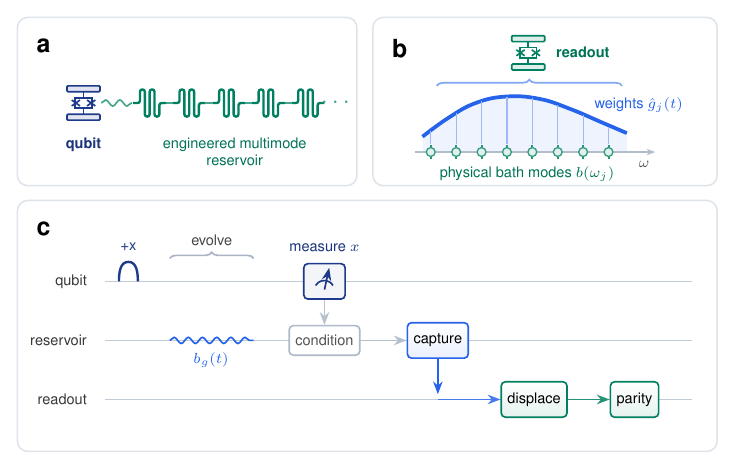}
\caption{\textbf{A circuit-QED protocol to measure the quantum record.} \textbf{(a)} A qubit strongly coupled ($\alpha_{\rm std}\approx1.6$) through $\sigma_z$ to an engineered multimode reservoir realizing a sub-Ohmic $J(\omega)$. The local resonator array realizes the physical reservoir sketched in Fig.~\ref{fig:concept}(c); the chain basis enters only as a simulation representation. \textbf{(b)} Capture in the physical frequency-mode basis: a readout element would couple with the time-dependent weights $\hat g_j(t)$ to the physical modes it collects and swap the collective mode $b_g(t)=\sum_j\hat g_j(t)b(\omega_j)$ into the measurement cavity; uncollected modes are traced out, which suppresses the fringes (Fig.~\ref{fig:thermal}). \textbf{(c)} Pulse sequence: prepare $\ket{+x}$, let the cat form for time $t$, measure $\sigma_x$ (outcome probabilities $P_{\pm x}=(1\pm\avg{\sigma_x})/2$), condition on the result, capture $b_g(t)$, then sweep displacements and photon-parity measurements to reconstruct $W^{(\pm x)}(\beta)$. The target Wigner signal is the conditional state illustrated in Fig.~\ref{fig:emergence}. Representative regime: $\Delta/2\pi\approx1$~GHz, for which $T/\omega_0=1$ corresponds to $\approx48$~mK. Capture of the time-dependent collective mode is the key experimental requirement.}
\label{fig:protocol}
\end{figure*}

Thermal noise weakens the recoverable quantum record, while the pointer record remains redundant in pure-dephasing fragments. To describe the physical warm bath, the thermalized chain represents it as an enlarged zero-temperature environment with positive- and negative-frequency modes~\cite{Tamascelli2019}. The physical laboratory mode is recovered through the thermofield (Bogoliubov) relation $b(\omega)=\sqrt{1+\nbar(\omega)}\,a_+(\omega)+\sqrt{\nbar(\omega)}\,a_-^\dagger(\omega)$, with $\nbar(\omega)=(e^{\omega/T}-1)^{-1}$ (Methods). We reconstruct the Wigner function of the physical thermal mode $b_g=\sum_j \hat g_j\,b(\omega_j)$, a canonical mode ($[b_g,b_g^\dagger]=1$) with thermal background occupation $\nbar_g=\sum_j\nbar(\omega_j)|\hat g_j|^2$. This quantity projects the initial thermal occupations onto the selected profile; it is not the total occupation of the driven mode.

In the pure-dephasing limit, each environmental branch evolves independently. The conditional state of the mode matched to the full branch displacement can then be solved exactly: it is a statistical mixture of cat states, with the two lobes corresponding to displaced thermal states. Supplementary Sec.~\ref{si:proofs} gives the density operator and proof.

\result{Thermal mixture theorem (Theorem 3)}{In pure dephasing, the physical mode matched to the full branch displacement has a conditional state that is exactly a thermal mixture of displaced cat states. Its normalized density operator and Wigner function are given in Supplementary Sec.~\ref{si:proofs}. Thermal mixing broadens both the displaced envelopes and the envelope of the interference term, without an additional uniform exponential damping factor multiplying that term.}

\noindent In pure dephasing the two branches carry opposite displacements
\begin{equation}
\beta_j(t)=\lambda_j\frac{e^{-i\Omega_jt}-1}{\Omega_j},\qquad
\hat g_j(t)=\frac{\overline{\beta_j(t)}}{\sqrt{\sum_k|\beta_k(t)|^2}},
\label{eq:response}
\end{equation}
and this matched mode underlies Theorem~3 and the exact fragment analysis. With tunneling we keep the occupation selector of the cold results and specify a physical detector band. The positive-frequency profile $g_j$ of the selector is filtered and renormalized,
\begin{equation}
g_j^{\rm det}=\frac{g_j\,\Theta(\Omega_j-0.1\omega_c)}{\sqrt{\sum_{\Omega_k\ge0.1\omega_c}|g_k|^2}},
\label{eq:detector}
\end{equation}
while every frequency retains its Bose occupation. For this occupation-derived profile in the sub-Ohmic bath, the unfiltered thermal background diverges at low frequency. The detector band makes that background finite by changing the measured mode; the bath state and its evolution are unchanged. The same selection procedure and detector band are used at all temperatures, including the vacuum baseline. The resulting mode weights can depend on both time and temperature. At each temperature and time, the mode is selected from the unconditional state, and the conditioned state is reconstructed in that mode.

With the common detector band $\omega\geq0.1\omega_c$, the sampled peak negative volumes are $0.448$ at $t\omega_0=0.64$ for the vacuum baseline and $0.204$ at $0.52$, $0.105$ at $0.46$, $0.044$ at $0.40$, and $0.0050$ at $0.28$ for $T/\omega_0=0.3$, $0.5$, $0.7$, and $1$, respectively (Fig.~\ref{fig:thermal}). Finite-temperature negativity is transient: the last sampled times above the $10^{-4}$ reporting threshold are $t\omega_0=1.30$, $1.06$, $0.70$, and $0.44$ at $T/\omega_0=0.3$, $0.5$, $0.7$, and $1$, respectively. The filtered vacuum baseline also decays, to $0.024$ at $t\omega_0=2.5$, whereas the unfiltered cold mode has $0.572$ (Fig.~\ref{fig:law}). The vacuum comparison demonstrates the effect of leaving the low-frequency band uncollected. In pure dephasing, thermal occupation of uncollected modes further suppresses interference [Supplementary Eq.~\eqref{eq:Sfragmentcontrast}]. The tunneling results reflect both incomplete collection and thermal effects, with mode weights also depending on temperature. Fringe suppression does not imply readable which-path information in the uncollected modes~\cite{Zwolak2009Mixed}.

The exact pure-dephasing matched mode, which collects every frequency, has $\Vnc=0.631$, $0.589$, $0.535$, $0.479$, and $0.404$ at $t\omega_0=2.5$ for $T/\omega_0=0$, $0.3$, $0.5$, $0.7$, and $1$ (Supplementary Sec.~\ref{si:proofs}). An unresolved negative volume in the detector band alone does not establish that the full conditional bath state is classical.

\section{Observing and retrieving the record}
The records are defined operationally by how they are read. The classical record is read by collecting physical frequency-band fragments. The quantum record is read by conditioning on a transverse qubit measurement and performing Wigner tomography of the selected physical collective bath mode. Structured bosonic environments, transverse qubit measurement, and single-mode phase-space readout have precedents in circuit and waveguide QED~\cite{Wallraff2004,Blais2021,Gu2017,Murch2013,Roy2017,Kannan2020}. Structured spectral densities can be synthesized with resonator arrays or frequency-dependent waveguide coupling~\cite{Chang2018,Kannan2020}. Targeting the sub-Ohmic form used here requires the corresponding low-frequency mode density and coupling profile. Figure~\ref{fig:protocol} sketches a possible realization and readout protocol.

\emph{Measurement principle.} For the chosen pure initial state, time evolution correlates the qubit pointer states $\ket{\!\uparrow},\ket{\!\downarrow}$ with branch states $\ket{\chi_\uparrow},\ket{\chi_\downarrow}$ of the entire bath [Fig.~\ref{fig:concept}(a)]. A transverse ($\sigma_x$) readout prepares a bath state proportional to $c_\uparrow\ket{\chi_\uparrow}\pm c_\downarrow\ket{\chi_\downarrow}$, with the amplitudes defined in Eq.~\eqref{eq:branch}. We reconstruct the reduced state of the selected collective mode, which is generally mixed even when the full conditional bath state is pure. The outcome probabilities are $P_{\pm x}=(1\pm\avg{\sigma_x})/2$; in the cold tunneling simulation $P_{+x}$ stays between about $0.36$ and $0.49$ after the first coherence minimum. We compute the $+x$ outcome. At finite temperature the preparation is described by the conditional density operator in Eq.~\eqref{eq:mixedcond}. For the pure branches, Theorem~1 provides a system--environment consistency check: the magnitude of the branch overlap equals $|\avg{\sigma_x}|$, while Wigner negativity separately certifies nonclassicality of the selected mode.

\emph{Mode capture and tomography.} The record is read from a \emph{time-dependent} physical collective mode $b_g(t)=\sum_j\hat g_j(t)b(\omega_j)$. At zero temperature it is represented in the chain basis by a linear combination $c_f(t)=\sum_k f_k(t)c_k$; at finite temperature its thermofield representation also contains creation operators (Methods). The key experimental step is therefore mode-matched capture. Retaining the late-time cold fringes requires collecting the low-frequency part of the mode: with a band $\omega\ge0.1\omega_c$, the negative volume decays after $t\omega_0\approx0.64$ (Fig.~\ref{fig:thermal}). A tunable coupler, or an ancilla resonator engineered with the appropriate profile, would swap this collective mode into a readout cavity [Fig.~\ref{fig:protocol}(b)]. Standard cavity Wigner tomography then reconstructs it: a calibrated phase-space displacement by $-\beta$, written $D(-\beta)$, followed by a photon-number-parity measurement through a dispersively coupled ancilla qubit~\cite{Brune1996} maps $W^{(\pm x)}(\beta)$ point by point~\cite{Banaszek1996,Lutterbach1997,Deleglise2008,Vlastakis2013} [Fig.~\ref{fig:protocol}(c)]. Figure~\ref{fig:emergence} shows the simulated signal for the cold mode. Negative Wigner values certify nonclassicality of the conditional state; comparison with the unconditional state quantifies the effect of conditioning.

\emph{Experimental requirements.} For an illustrative scale $\omega_0/2\pi\approx1$~GHz and $\Delta=\omega_0$, the simulated range $0\le T/\omega_0\le1$ corresponds to $0\le T_{\rm phys}\lesssim48$~mK, and $t\omega_0\le2.5$ corresponds to about $0.4$~ns. Capturing the distributed, time-dependent mode requires control of its weights and phases, sufficient collection efficiency, and a suitably short transfer duration, in addition to qubit readout and cavity tomography. Quantitative device design and capture fidelities remain to be established.

\section{Discussion}
The results make the environment a measurable part of a decoherence experiment. In pure dephasing, the same system--bath state supports both a conditional nonclassical collective mode and redundant pointer information in physical frequency-band fragments. With tunneling, the reconstruction establishes the conditional nonclassical state and its time evolution. For the pure initial states considered here, the coherence--overlap identity links the residual qubit coherence to the distinguishability of the full bath branches. Fragment information separately describes how the pointer record is distributed.

A transverse qubit measurement reveals additional structure in a selected collective bath mode: a conditional state with stronger Wigner negativity than its unconditional counterpart. This nonclassical state and the distributed which-path record require different measurements of the same bath. At zero temperature, the nonclassical record in the unfiltered mode persists over the simulated interval. In the pure-dephasing limit, finite temperature yields an exact thermal mixture of displaced cat states. With tunneling, a detector band that excludes frequencies below $0.1\omega_c$ sees peak negativities $0.448$, $0.204$, $0.105$, $0.044$, and $0.0050$ at $T/\omega_0=0$, $0.3$, $0.5$, $0.7$, and $1$; at finite temperature the negativity is transient. Its suppression reflects both uncollected modes and thermal effects, and it does not imply readable which-path records in those modes~\cite{Zwolak2009Mixed}. Separately, pure-dephasing information bounds quantify the classical pointer information in frequency-band fragments described by quantum Darwinism~\cite{Zurek2003,OllivierPoulinZurek2004,Zurek2009Darwinism}. Together, collective-mode tomography and fragment measurements would allow the two forms of environmental record to be compared in the same device.

Quantum Darwinism has been analyzed in an oscillator bath~\cite{BlumeKohout2008}, in spin and structured environments~\cite{Zwolak2016,Giorgi2015,Ryan2021}, in disordered many-body environments~\cite{Mirkin2021}, in environments whose internal interactions make redundancy rise and fall~\cite{Riedel2012}, and in mixed environments whose initial entropy reduces record capacity~\cite{Zwolak2009Mixed}. Here the physical frequency modes are noninteracting; the chain couplings arise from a change of basis. Objectivity requires more than a partial-information plateau: it has been tested through spectrum broadcast structures~\cite{Kwiatkowski2021,Korbicz2021}, while studies of unequal fragments emphasize the role of fragment averaging~\cite{Chisholm2023,Chisholm2024}. Redundant local records also constrain how a many-body state decomposes into classical branches~\cite{Riedel2017}. Accessible classical information and quantum discord are complementary~\cite{ZwolakZurek2013}. Conditional preparation of a cavity cat by measuring a probe atom is also established~\cite{Brune1992}. Our contribution is the reservoir-side reconstruction: in a structured sub-Ohmic bath with tunneling, we reconstruct the conditional phase-space state of a collective bath mode, follow its formation and its suppression through a specified detector band, and compare it with exact fragment records in the pure-dephasing limit.

\section{Methods}
\textit{Model and conventions.} We use $\omega_0=\omega_c/4=1$ as the fixed frequency unit, with $\Delta=\omega_0$ for tunneling and $\Delta=0$ for pure dephasing. The isolated qubit gap is $2\Delta$. The spectral density Eq.~\eqref{eq:Jw} uses $\sigma_z$ coupling; matching to $J_{\rm std}=2\pi\alpha_{\rm std}\omega_c^{1-s}\omega^s e^{-\omega/\omega_c}$ with $(\sigma_z/2)$ coupling gives $\alpha_{\rm std}=4\alpha$. Temperature is $T/\omega_0=k_BT_{\rm phys}/(\hbar\omega_0)$.

\textit{Chain mapping and thermalized bath.} We generate the chain coefficients $(\omega_n,t_n,g)$ by a stable Lanczos recurrence on a graded Gauss--Legendre discretization of the (thermalized) spectral measure; the zero-temperature support is $[0,\omega_{\max}]$ and the finite-temperature support extends to $[-\omega_{\max},\omega_{\max}]$ following T-TEDOPA~\cite{Tamascelli2019}, with $\omega_{\max}=8\omega_c$. The chain length is chosen using a light-cone estimate to suppress boundary reflections over the simulated window, with a minimum of $N=120$ sites. The chain lengths and boundary occupations are listed in Supplementary Table~\ref{tab:Sconv}. The chain is initialized in vacuum.

\textit{Time evolution.} We evolve a matrix-product-state (MPS) representation of the qubit and chain with two-site TDVP~\cite{Haegeman2011,Haegeman2016,Paeckel2019}. A shifted boson basis centers the real mean amplitude of each bath site before every step and restores the original basis afterwards. We use a time step $\Delta t\omega_0=0.02$, a singular-value cutoff of $10^{-10}$, and a Krylov tolerance of $10^{-9}$. Local Fock dimensions are chosen from the displacement scale, with a minimum dimension of $18$ on every site.

\textit{Wigner reconstruction.} We evaluate the collective characteristic function as an MPS overlap with a product of single-site displacement operators and then Fourier transform it~\cite{Vogel1989,Lvovsky2009}. For $b_g=\sum_k f_kc_k$, the amplitudes are $\eta\overline f_k$. We use exact oscillator displacement matrix elements projected onto the retained Fock basis. The characteristic-function domain is enlarged until $|\chi|<10^{-7}$ at eight sampled outer-domain points, and no window function is applied. Reconstructions on finer grids at the peak, a later time, and the endpoint of each conditional time series test the discretization error. The zero-temperature vacuum has zero negativity analytically. Supplementary Sec.~\ref{si:wigner} compares the reconstructions with exact results and gives their accuracy.

\textit{Conditioning and the physical thermal mode.} The collective mode is selected from the unconditional state and used unchanged for the $+x$-conditioned Wigner function. In the thermofield basis, $D_g(\eta)$ has amplitudes $\mu(+\Omega_j)=\eta\overline u_j$ and $\mu(-\Omega_j)=-\bar\eta v_j$, with $u_j=\sqrt{1+\nbar_j}\hat g_j$ and $v_j=\sqrt{\nbar_j}\hat g_j$. Projection onto the finite chain is supplemented by the analytic Gaussian factor of the omitted vacuum directions. Canonicality follows from $\sum_j|\hat g_j|^2=1$. All occupations retain $\nbar_j=(e^{\Omega_j/T}-1)^{-1}$. The temperature comparison filters the selector weights with Eq.~\eqref{eq:detector}; the cold results use no filter. Wigner functions are reconstructed at all $126$ time steps of each trajectory. Proofs and numerical details are given in the Supplementary Information, which includes Refs.~\cite{JeongRalph2006,Szego1939,Scutaru1998,Banchi2015,Holevo1973,Uhlmann1976,Zwolak2010}.

\section*{Acknowledgments}
The authors acknowledge the use of the ABCI-Q System H~\cite{GQuAT_ABCIQ_2026} at G-QuAT, AIST, and the Miyabi-G system~\cite{JCAHPC_Miyabi_2026} at JCAHPC for numerical simulations. This work was partially supported by the Support Center for Advanced Telecommunications Technology Research and the Murata Science Foundation.

\section*{Data and code availability}
The simulation and analysis code, together with the numerical data underlying the figures, are available in the public repository at \url{https://github.com/shrakashlab/QuantumReservoirTomography}. Additional tensor-network states and Wigner grids are available from the authors upon reasonable request.

\section*{Author contributions}
J.A. conceived the study, developed the simulations and analysis, prepared the figures, and wrote the manuscript. K.A., T.K., and H.N. contributed to data generation, checking, and figure evaluation. D.T. and M.S. contributed to discussions and interpretation. J.S. supervised the project. All authors reviewed the manuscript.

\section*{Competing interests}
The authors declare no competing interests.

\putbib[reference]
\end{bibunit}

\clearpage
\onecolumngrid
\begin{center}
{\large\textbf{Supplementary Information}}\\[2pt] \textit{Complementary quantum and classical records of qubit decoherence}
\end{center}
\vspace{6pt}
\twocolumngrid

\setcounter{figure}{0}\renewcommand{\thefigure}{S\arabic{figure}}
\setcounter{equation}{0}\renewcommand{\theequation}{S\arabic{equation}}
\setcounter{table}{0}\renewcommand{\thetable}{S\arabic{table}}
\setcounter{section}{0}\renewcommand{\thesection}{S\arabic{section}}
\renewcommand{\theHfigure}{S\arabic{figure}}
\renewcommand{\theHequation}{S\arabic{equation}}
\renewcommand{\theHtable}{S\arabic{table}}
\renewcommand{\theHsection}{S\arabic{section}}

\begin{bibunit}[apsrev4-2]
\renewcommand{\citenumfont}[1]{S#1}
\renewcommand{\bibnumfmt}[1]{[S#1]}

\makeatletter
\def\hyper@natlinkstart#1{%
  \Hy@backout{#1}%
  \hyper@linkstart{cite}{cite.SI.#1}%
  \def\hyper@nat@current{#1}%
}
\def\hyper@natlinkend{\hyper@linkend}
\def\hyper@natlinkbreak#1#2{%
  \hyper@linkend#1\hyper@linkstart{cite}{cite.SI.#2}%
}
\def\hyper@natanchorstart#1{%
  \Hy@raisedlink{\hyper@anchorstart{cite.SI.#1}}%
}
\def\hyper@natanchorend{\hyper@anchorend}
\makeatother

\section{Model, conventions, and units}
\label{si:model}
The spectral density Eq.~\eqref{eq:Jw} defines $\alpha$; numerical values must be translated before comparison with conventions that absorb factors of $\pi$. Matching our $\sigma_z$ coupling to the standard form $J_{\rm std}=2\pi\alpha_{\rm std}(\cdots)$ for $(\sigma_z/2)$ coupling gives $\alpha_{\rm std}=4\alpha$. The isolated qubit gap is $2\Delta$. Temperature is reported as $T/\omega_0=k_BT_{\rm phys}/(\hbar\omega_0)$, where $\omega_0=\omega_c/4$. For the tunneling simulations $\Delta=\omega_0$, so the following conversions can equivalently use $\Delta$. Writing $f_\Delta=\Delta/2\pi$ gives the qubit transition frequency $f_q=2f_\Delta$ and, at $T/\omega_0=1$, $T_{\rm phys}=hf_\Delta/k_B=hf_q/(2k_B)$. Thus $f_\Delta=1$~GHz (a $2$~GHz qubit transition) corresponds to $48$~mK; equivalently, a $1$~GHz qubit transition corresponds to $24$~mK. The Bose occupation at the gap at $T/\omega_0=1$ is $\nbar(2\Delta)\approx0.16$, a moderate thermal regime below the cutoff $\omega_c=4\Delta$.

\section{Exact results: proofs of Theorems 1--3}
\label{si:proofs}

\subsection{Theorem 1 (coherence--overlap identity)}
The unbiased Hamiltonian $H=\Delta\sigma_x+\sigma_z B+H_B$, with $B=\sum_k\lambda_k(b_k+b_k^\dagger)$ and $H_B=\sum_k\Omega_kb_k^\dagger b_k$, commutes with $\Pi=\sigma_x\otimes P$, where $Pb_kP=-b_k$. Both signs in $\sigma_zB$ reverse under $\Pi$, leaving $H$ invariant. Starting from $\ket{+x}\otimes\ket{\phi_B}$ with $P\ket{\phi_B}=\ket{\phi_B}$, the pure global state therefore satisfies $\Pi\ket{\Psi(t)}=\ket{\Psi(t)}$. A parity-compatible thermofield purification obeys the same pure-state statement in the enlarged space.

Expanding in the pointer ($\sigma_z$) basis, $\ket{\Psi}=c_\uparrow\ket{\uparrow}\ket{\chi_\uparrow}+c_\downarrow\ket{\downarrow}\ket{\chi_\downarrow}$ with normalized $\ket{\chi_{\uparrow,\downarrow}}$, invariance $\Pi\ket{\Psi}=\ket{\Psi}$ gives $c_\uparrow\ket{\chi_\uparrow}=c_\downarrow P\ket{\chi_\downarrow}$. Taking norms, $|c_\uparrow|=|c_\downarrow|=1/\sqrt2$. Writing $c_\downarrow=c_\uparrow e^{i\varphi}$ gives $\ket{\chi_\uparrow}=e^{i\varphi}P\ket{\chi_\downarrow}$, so $\braket{\chi_\uparrow}{\chi_\downarrow}=e^{-i\varphi}\bra{\chi_\downarrow}P\ket{\chi_\downarrow}$ with $\bra{\chi_\downarrow}P\ket{\chi_\downarrow}\in\mathbb{R}$ (as $P=P^\dagger$). Substituting into $\avg{\sigma_x}=2\,\mathrm{Re}[\bar c_\uparrow c_\downarrow\braket{\chi_\uparrow}{\chi_\downarrow}]$ and using $\bar c_\uparrow c_\downarrow=\tfrac12 e^{i\varphi}$, the phases cancel: $\avg{\sigma_x}=\bra{\chi_\downarrow}P\ket{\chi_\downarrow}\in\mathbb{R}$, while $|\braket{\chi_\uparrow}{\chi_\downarrow}|=|\bra{\chi_\downarrow}P\ket{\chi_\downarrow}|$. Therefore $|\avg{\sigma_x}(t)|=|\braket{\chi_\uparrow}{\chi_\downarrow}(t)|$ at all $t$, $\Delta$, and coupling for these pure branches; at finite temperature these are purified branches.\par\smallskip\noindent The conditional $+x$ state is $\propto\ket{\chi_\uparrow}+e^{i\varphi}\ket{\chi_\downarrow}$. The overlap fixes the reduced qubit coherence and the normalization of this conditional state; its phase-space negativity is a separate functional of the two branch states and is evaluated directly from the Wigner function.

For the physical mixed thermal state, parity invariance is $\Pi\rho_{SB}\Pi=\rho_{SB}$. It implies $\avg{\sigma_y}=\avg{\sigma_z}=0$ and $V=2|\Tr_B X_B|$, with $X_B=\bra{\uparrow}\rho_{SB}\ket{\downarrow}$. This is distinct from the root fidelity of the two normalized physical bath branches.

\subsection{Theorem 2 (visibility--distinguishability relation)}
Absorb the phases of $c_\uparrow$ and $c_\downarrow$ into the normalized branch kets. The balanced global state is then $\ket{\Psi}=\tfrac{1}{\sqrt2}(\ket{\uparrow}\ket{\chi_\uparrow} +\ket{\downarrow}\ket{\chi_\downarrow})$. Its reduced qubit state has Bloch vector $\bm r=(\avg{\sigma_x},0,0)$ with $\avg{\sigma_x}=\mathrm{Re}\braket{\chi_\uparrow}{\chi_\downarrow}$ (Theorem~1), so $|\bm r|=V=|\avg{\sigma_x}|$ and the purity is $\Tr\rho_S^2=(1+V^2)/2$. The environment's ability to distinguish the two pointer branches is their trace distance, $D=\tfrac12\|\rho_B^\uparrow-\rho_B^\downarrow\|_1$. For pure branches, $D=\sqrt{1-|\braket{\chi_\uparrow}{\chi_\downarrow}|^2}=\sqrt{1-V^2}$, equivalently $D^2=2(1-\Tr\rho_S^2)$, the normalized linear entropy of the qubit. Hence $V^2+D^2=1$.\par\smallskip\noindent This is the Englert visibility--distinguishability duality~\cite{WoottersZurek1979,Jaeger1995,Englert1996} realized exactly along the decoherence trajectory. Here it provides a system--environment consistency relation for a strongly coupled structured bath: $V$ is the coherent branch overlap and $D$ the which-path distinguishability of the complete bath. Full-bath distinguishability does not determine how that information is distributed over fragments; redundancy is a separate question~\cite{OllivierPoulinZurek2004,Zurek2009Darwinism,Korbicz2021}, examined in Sec.~\ref{si:darwin}. At $T>0$ the duality holds in the thermofield purification with pure branch kets; for the \emph{physical} mixed branches it relaxes to the inequality $V^2+D^2\le1$, and the pure-dephasing information bounds below show reduced physical distinguishability with thermal noise.

\subsection{Theorem 3 (thermal mixture of displaced cats)}
At $\Delta=0$, the two forced-oscillator propagators have opposite displacements $\pm\bm\beta(t)$ from Eq.~\eqref{eq:response} and a common scalar phase. The unnormalized physical conditional bath operator is $K\rho_B^{\rm th}K^\dagger/4$, where $K=D(\bm\beta)+D(-\bm\beta)$. For $B^2=\sum_j|\beta_j|^2>0$, choose $\hat g_j=\overline\beta_j/B$. Then $D(\bm\beta)=D_g(B)$ and the marginal of $g$ is thermal with $\nbar_g=\sum_j\nbar_j|\beta_j|^2/B^2$. A rotation can correlate thermal modes with unequal occupations, but $K$ acts only on $g$, so its marginal calculation is exact:
\begin{equation}
\rho_g^{(+x)}=\frac{K_g\rho_{\rm th}(\nbar_g)K_g^\dagger}{Z},\quad
K_g=D_g(B)+D_g(-B),
\label{eq:Smix}
\end{equation}
where $Z=2[1+e^{-2s_gB^2}]$ and $s_g=2\nbar_g+1$. A thermal coherent-state decomposition gives a mixture of the unnormalized cats $K_g\ket{\gamma}$; normalized component cats are weighted by their outcome probabilities. The physical conditional mode is generally mixed.

Using $\Tr[\rho_{\rm th}D(\eta)]=e^{-s_g|\eta|^2/2}$ gives
\begin{align}
\chi_g(\eta)=\frac{1}{Z}\big[&2e^{-s_g|\eta|^2/2}\cos(2B\,\mathrm{Im}\eta)\nonumber\\
&+e^{-s_g|\eta-2B|^2/2}+e^{-s_g|\eta+2B|^2/2}\big].
\label{eq:Schar}
\end{align}
For $G_s(z)=2e^{-2|z|^2/s}/(\pi s)$, its Fourier transform is
\begin{align}
W_g^{(+x)}(z)=\frac{1}{Z}\big[&G_{s_g}(z-B)+G_{s_g}(z+B)\nonumber\\
&+2G_{s_g}(z)\cos(4B\,\mathrm{Im}z)\big].
\label{eq:Swigner}
\end{align}
The thermal visibility $V_T=e^{-2s_gB^2}$ fixes normalization and the qubit outcome probability; it is not an additional uniform multiplier of the Wigner interference term. Conditional preparation from displaced thermal components has established quantum-optical precedents~\cite{JeongRalph2006}; the present formula specifies the reservoir-matched mode and its continuum occupation.

\subsection{Negativity at \texorpdfstring{$\Delta=0$}{Delta=0}}
At $T=0$, Eq.~\eqref{eq:Smix} reduces to an even cat. At finite $T$, negativity follows by integrating Eq.~\eqref{eq:Swigner}. For finite $s_g$ and $B>0$, evaluating at $\mathrm{Re}\,z=0$, $\mathrm{Im}\,z=\pi/(4B)$ gives a negative factor $e^{-2B^2/s_g}-1$. This proves negativity somewhere for the ideal full matched mode, without guaranteeing experimental resolution at arbitrary temperature. At $t\omega_0=2.5$, $B=2.768$ and $(\nbar_g,\Vnc,\Tr\rho_g^2)$ are $(0,0.631,1)$, $(0.446,0.589,0.529)$, $(0.876,0.535,0.363)$, $(1.334,0.479,0.273)$, and $(2.043,0.404,0.197)$ for $T/\omega_0=0,0.3,0.5,0.7,1$. When the number of frequency nodes is doubled or the upper frequency limit is raised from $8\omega_c$ to $12\omega_c$, $B$ and $\nbar_g$ change by less than $4\times10^{-6}$, and $\Vnc$ and the purity by less than $9\times10^{-7}$.

For the displacement-matched mode within a physical frequency fragment $F$, define $B_F^2=\sum_{j\in F}|\beta_j|^2>0$ and $s_F=1+2\sum_{j\in F}\nbar_j|\beta_j|^2/B_F^2$. Replace $B$ and $s_g$ in Eq.~\eqref{eq:Swigner} by $B_F$ and $s_F$, and multiply the interference term by
\begin{equation}
r_{\bar F}=\exp[-2\!\sum_{j\notin F}(2\nbar_j+1)|\beta_j|^2].
\label{eq:Sfragmentcontrast}
\end{equation}
Its normalization is $2[1+r_{\bar F}e^{-2s_FB_F^2}]$. Thus uncollected modes multiply the interference term by a uniform factor, unlike thermal broadening of the full mode.

\section{Chain mapping}
\label{si:chain}
The (thermalized) measure is discretized on a graded composite Gauss--Legendre grid (geometric refinement near $\omega=0$), and chain coefficients are the three-term recurrence coefficients of the resulting orthogonal-polynomial measure~\cite{Szego1939}, computed by Lanczos with full reorthogonalization. For $T>0$ the nearly symmetric measure on the $\pm\omega$ support produces small alternating on-site energies in the chain tail. The chain length is chosen to keep boundary occupation negligible over the simulated window.

\section{Thermalized representation and the physical thermal mode}
\label{si:thermo}
T-TEDOPA represents the thermal bath as an extended zero-temperature environment with modes on $\pm\omega$~\cite{Tamascelli2019}. The physical mode is recovered by the thermofield Bogoliubov relation $b(\omega)=\sqrt{1+\nbar}\,a_+(\omega)+\sqrt{\nbar}\,a_-^\dagger(\omega)$. For the collective mode $b_g=\sum_j\hat g_j b(\omega_j)$, the displacement $D(\eta)=\exp(\eta b_g^\dagger-\bar\eta b_g)$ has per-node amplitudes $\mu(+\omega_j)=\eta\,\overline{u_j}$ and $\mu(-\omega_j)=-\bar\eta\,v_j$ with $u_j=\sqrt{1+\nbar_j}\,\hat g_j$, $v_j=\sqrt{\nbar_j}\,\hat g_j$. Projecting onto the chain modes gives a two-amplitude characteristic function $\chi(\eta)=\avg{\prod_k\exp(\mu_k^{\rm ch}c_k^\dagger-\mathrm{h.c.})}$, multiplied by the analytic Gaussian factor of the complement weight that lies outside the chain span. Here $\mu_k^{\rm ch}=\eta\,\overline{f_{u,k}}+\bar\eta\,\overline{f_{v,k}}$, where $f_{u,k}$ and $f_{v,k}$ are the chain components of $u_j$ on the $+\omega_j$ nodes and of $-\bar v_j$ on the $-\omega_j$ nodes. The mode is canonical, $[b_g,b_g^\dagger]=\sum_j(|u_j|^2-|v_j|^2)=1$, and the thermal background occupation is $\nbar_g=\sum_j\nbar_j|\hat g_j|^2$.

\emph{Infrared behavior and mode choice.} For the matched mode of Eq.~\eqref{eq:response}, $|\beta(\omega)|^2\sim\omega^s$ and $\nbar(\omega)|\beta(\omega)|^2\sim\omega^{s-1}$ at finite time, so its thermal background is finite for $s>0$. The occupation-derived profile of the thermalized chain instead weights low frequencies by an extra factor $1+\nbar(\omega)$, so for $s=0.5$ its thermal background diverges as the lower frequency limit approaches zero. Replacing low-frequency occupations by zero would mix physical bath and purification operators and change the observable. The detector band of Eq.~\eqref{eq:detector} instead removes those frequencies from the measured mode and leaves every occupation unchanged; Fig.~\ref{fig:thermal} uses it at all five temperatures and shows the thermal background $\nbar_g(t)$ of the detected mode.

\section{Tensor-network method}
\label{si:tn}
Two-site TDVP uses a shifted boson basis that centers the mean oscillator coordinate $\avg{c_k+c_k^\dagger}=2\,\mathrm{Re}\avg{c_k}$ before each time step and restores the original basis afterwards. The simulations use at least $120$ chain sites (more when the light-cone criterion requires), $D_{\max}=400$, $\Delta t\omega_0=0.02$, singular-value cutoff $10^{-10}$, Krylov dimension $12$, tolerance $10^{-9}$, and a minimum local Fock dimension of $18$. Table~\ref{tab:Sconv} lists the resulting chain lengths, bond dimensions, and accuracy diagnostics; convergence tests are given in Sec.~\ref{si:conv}.

\begin{table*}[t]
\caption{Diagnostics of the tunneling trajectories ($\Delta=\omega_0$, $s=0.5$, $\alpha=0.4$). Maxima are taken over $0\le t\omega_0\le2.5$. Fock edge is the largest population of the highest retained number state on any bath site, boundary occupation is summed over the last $20$ chain sites, and the Wigner normalization refers to the detector-band mode.}
\label{tab:Sconv}
\centering
\begin{ruledtabular}
\begin{tabular}{ccccccc}
$T/\omega_0$ & chain length & max bond & Fock edge & max $|\Delta E|/\omega_0$ & max $|\int W-1|$ & boundary occupation \\
$0$ & $120$ & $18$ & $7.3\times10^{-12}$ & $3.8\times10^{-6}$ & $1.0\times10^{-10}$ & $1.1\times10^{-25}$ \\
$0.3$ & $120$ & $30$ & $1.9\times10^{-9}$ & $5.7\times10^{-6}$ & $1.5\times10^{-10}$ & $3.6\times10^{-25}$ \\
$0.5$ & $120$ & $39$ & $5.5\times10^{-7}$ & $7.2\times10^{-6}$ & $4.7\times10^{-11}$ & $4.0\times10^{-25}$ \\
$0.7$ & $120$ & $54$ & $3.8\times10^{-5}$ & $7.0\times10^{-6}$ & $4.1\times10^{-13}$ & $3.8\times10^{-25}$ \\
$1$ & $120$ & $87$ & $4.6\times10^{-4}$ & $5.6\times10^{-6}$ & $9.5\times10^{-12}$ & $8.9\times10^{-26}$ \\
\end{tabular}
\end{ruledtabular}
\end{table*}

\section{Wigner reconstruction}
\label{si:wigner}
The collective characteristic function is evaluated as an MPS overlap and Fourier transformed. Each oscillator displacement is $P_dD(\eta)P_d$, where $P_d$ projects onto the retained $d$ Fock states; its matrix elements are evaluated using associated Laguerre polynomials. Exponentiating the truncated generator, $\exp[P_d(\eta a^\dagger-\bar\eta a)P_d]$, can instead produce artificial large-$\eta$ revivals even in vacuum. The characteristic-function domain is enlarged until $|\chi|<10^{-7}$ at eight sampled outer-domain points.

All reconstructed Wigner functions are normalized to within $1.6\times10^{-10}$, and $|\chi|$ is at most $9.6\times10^{-8}$ at these sampled points. Repeating the reconstruction on $1.5$ times denser grids, with the domain enlarged until $|\chi|<10^{-9}$ at the same points, at the peak, a later time, and the endpoint of each conditional time series, changes $\Vnc$ by at most $3.4\times10^{-4}$; the quoted peak and final values use these finer grids. The largest difference between $|\avg{\sigma_x}|$ and the independently evaluated branch overlap in Fig.~\ref{fig:law} is $2.4\times10^{-15}$. At $T=0$, a pure-dephasing evolution reconstructed with the exponentiated truncated generator agrees with the exact negative volume within $7.2\times10^{-4}$ for $t\omega_0\ge0.2$.

At $t\omega_0=2.5$, the cold conditional mode has purity $\Tr\rho_g^2=0.870$ and even-cat fidelity $F_{\rm cat}=\bra{C_+(\beta)}\rho_g\ket{C_+(\beta)}=0.930$, where $\ket{C_+(\beta)}=(\ket{\beta}+\ket{-\beta})/\sqrt{2(1+e^{-2|\beta|^2})}$. The fit varies the amplitude and orientation of $\beta$ over centered even and odd cats; the best fit is even.

\section{Mode definitions and conditioning}
\label{si:mode}
The bath one-body density matrix is $C_{mn}=\avg{c_m^\dagger c_n}$. Its most highly occupied eigenvector defines the leading natural orbital, $c_f(t)=\sum_k f_k(t)c_k$. Because an eigenvector has an arbitrary complex phase, we align its phase with the orbital from the preceding time step, removing arbitrary phase jumps. At $t=0$, where the vacuum occupation matrix is degenerate, we use the first chain mode $c_0$, which couples to the qubit. For the thermal comparison, its positive-frequency image is filtered and normalized with Eq.~\eqref{eq:detector}, and the initial Wigner function is the analytical thermal Gaussian. At finite temperature the selector is derived from the positive-frequency sector of the thermalized chain. Its occupation criterion refers to this auxiliary representation, so it does not maximize the occupation of the physical thermal bath. It defines a selection rule for the measured physical coordinate, with the detector band applied at all temperatures.

At each time the selector is the leading orbital of the unconditional state, unfiltered for the cold results and filtered by Eq.~\eqref{eq:detector} for the temperature comparison, and it is used unchanged after conditioning. Both Wigner functions are reconstructed for the cold mode (Figs.~\ref{fig:emergence} and~\ref{fig:law}); the temperature comparison reconstructs the conditioned state. The unconditional state is obtained after tracing out the qubit and therefore mixes the two pointer branches. The conditional state is obtained by projecting the qubit onto the $+x$ result and normalizing by its probability, as defined in Eq.~\eqref{eq:cond} for a pure global state and Eq.~\eqref{eq:mixedcond} for a mixed state. Comparing the two states in the same mode isolates the effect of the measurement; otherwise a change of mode could be mistaken for a change of quantum state. The conditional negativity is therefore a record selected from the joint system--bath state.

\begin{figure*}[!t]
\centering
\includegraphics[width=0.88\textwidth]{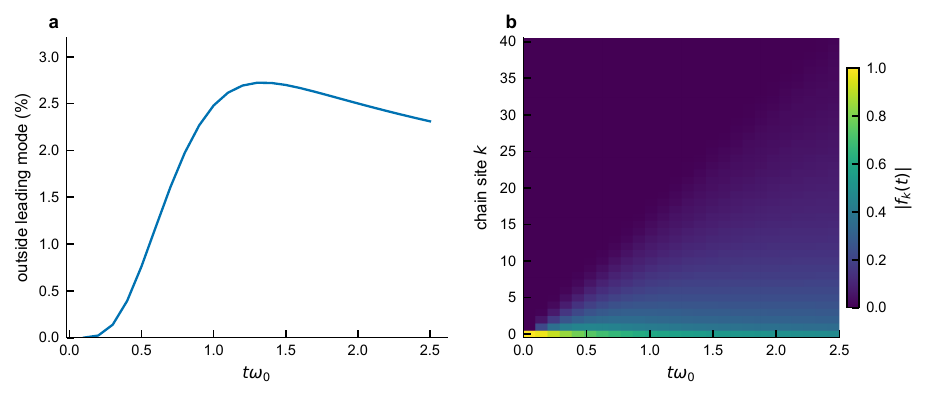}
\caption{\textbf{Mode structure.} \textbf{(a)} Residual occupation outside the leading natural orbital, $100(1-\lambda_1/\sum_j\lambda_j)$; it stays below $3\%$, equivalently $\lambda_1/\sum_j\lambda_j>0.97$. \textbf{(b)} Chain-basis profile $|f_k(t)|$ of the same leading orbital as it propagates along the mapped chain ($s=0.5$, $T=0$, $\alpha=0.4$).}
\label{fig:Smode}
\end{figure*}

Figure~\ref{fig:Smode} reports occupation outside the leading orbital and its chain profile.

\section{Quantum Darwinism: concentration vs.\ redundancy}
\label{si:darwin}
The conditional single-mode state and the physical fragment information are different observables. The leading cold orbital carries about $97\%$ of the one-body occupation, but this fraction is not a bound on the information outside that mode. Frequency-band fragments are defined in the original star bath, not by neighboring chain sites.

\begin{figure*}[!t]
\centering
\includegraphics[width=\textwidth]{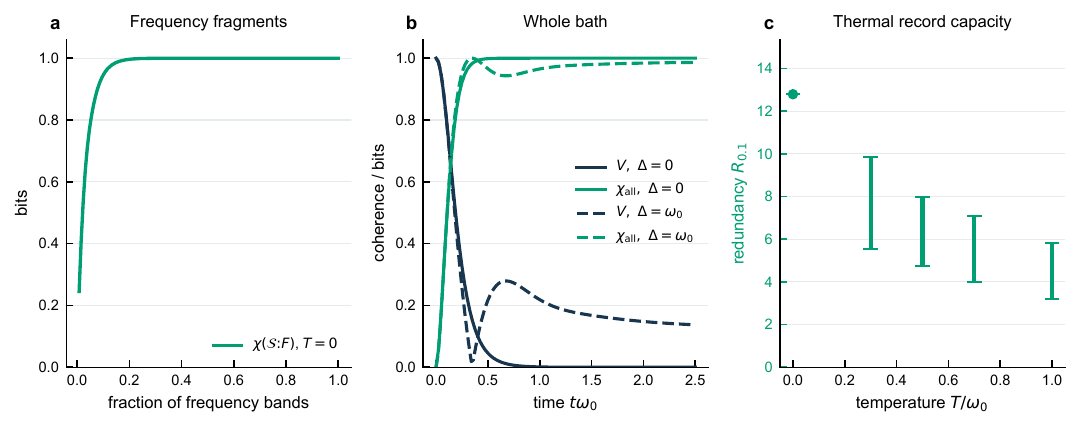}
\caption{\textbf{Redundant classical records in pure dephasing.} \textbf{(a)} At $\Delta=0$, $T=0$, $t\omega_0=2.5$, the mean Holevo information of random physical frequency-band fragments rises rapidly toward one bit; shading gives sampling standard errors. \textbf{(b)} Qubit coherence and whole-bath Holevo information for pure cold branches, in pure dephasing (solid) and with tunneling (dashed). \textbf{(c)} Sampled redundancy versus temperature at $\Delta=0$; interval widths arise from the information bounds.}
\label{fig:Sdarwin}
\end{figure*}

\begin{figure*}[!t]
\centering
\includegraphics[width=\textwidth]{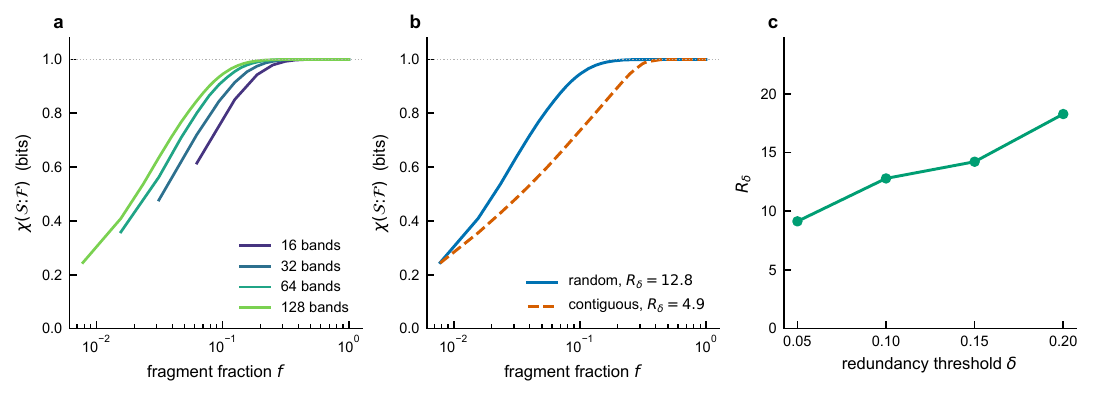}
\caption{\textbf{Partition dependence of the Darwinian estimate} ($\Delta=0$, $T=0$, $t\omega_0=2.5$, $s=0.5$, $\alpha=0.4$). \textbf{(a)} Mean partial-information curves for $16$, $32$, $64$, and $128$ physical frequency bands. \textbf{(b)} Random and contiguous fragments both exhibit plateaus but give different redundancy estimates. \textbf{(c)} The estimated redundancy changes with threshold $\delta$.}
\label{fig:Sdarwinrobust}
\end{figure*}

For pure dephasing the branches are products of displaced thermal states. With $\beta_j$ from Eq.~\eqref{eq:response}, define
\begin{equation}
w_F=4\sum_{j\in F}\frac{|\beta_j|^2}{2\nbar_j+1},\qquad
F_r(F)=e^{-w_F/2}.
\end{equation}
Here $F_r=\|\sqrt{\rho_F^\uparrow}\sqrt{\rho_F^\downarrow}\|_1$ is the root fidelity, which for displaced thermal states takes this closed form~\cite{Scutaru1998,Banchi2015}. The Holevo information~\cite{Holevo1973} is
\begin{equation}
\chi_F=S[(\rho_F^\uparrow+\rho_F^\downarrow)/2]
-[S(\rho_F^\uparrow)+S(\rho_F^\downarrow)]/2,
\end{equation}
in bits. For pure cold branches it equals $U_F=h_2[(1+e^{-w_F/2})/2]$. For thermal branches $\chi_F\le U_F$, from optimal purifications and data processing~\cite{Holevo1973,Uhlmann1976}. Measuring the displacement-aligned quadratures gives a sufficient statistic distributed as $\mathcal N(\pm\sqrt{w_F},1)$ and a lower bound
\begin{equation}
L_F=1-\mathbb E_{z\sim\mathcal N(0,1)}\log_2[1+e^{-2w_F-2\sqrt{w_F}z}].
\end{equation}
Thus $L_F\le\chi_F\le U_F$. These Gaussian expressions do not apply to the generally non-Gaussian branches at $\Delta\ne0$.

Frequency integrals use $96$ Gauss--Legendre nodes per panel, a graded grid reaching zero, and upper limit $8\omega_c$. Integration panels are aligned with the boundaries of the $128$ logarithmic bands; the first band includes all frequencies below $0.01\omega_c$. We use $4096$ random permutations to form nested fragments of every size. Standard errors quantify sampling of the mean curves.

For the common target $(1-\delta)\chi_{\rm all}$, where $\chi_{\rm all}$ is the Holevo information of all $M$ bands, let $m_{\rm possible}$ first satisfy $\overline U(m)\ge(1-\delta)L_{\rm all}$ and $m_{\rm certain}$ first satisfy $\overline L(m)\ge(1-\delta)U_{\rm all}$. The mean-curve redundancy interval is $M/m_{\rm certain}\le R_\delta\le M/m_{\rm possible}$. Its endpoints carry additional sampling uncertainty, which we estimate from sixteen independent batches of $256$ permutations.

At $t\omega_0=2.5$ and $\delta=0.1$, the sampled estimates are $R_\delta=12.80$ at $T=0$ and $R_\delta\in[5.57,9.85]$, $[4.74,8.00]$, $[4.00,7.11]$, and $[3.20,5.82]$ at $T/\omega_0=0.3$, $0.5$, $0.7$, and $1$. The cold batches range from $11.64$ to $12.80$; this sampling variation is separate from the finite-temperature information-bound intervals.

The full physical thermal visibility and root fidelity differ:
\begin{align}
V_T&=e^{-2\sum_j(2\nbar_j+1)|\beta_j|^2},\\
F_r(\mathrm{bath})&=e^{-2\sum_j|\beta_j|^2/(2\nbar_j+1)}.
\end{align}
Heating can therefore dephase the qubit more efficiently while making the physical branches less distinguishable. The decrease of both redundancy bounds with temperature is consistent with reduced record capacity in mixed environments~\cite{Zwolak2009Mixed,Zwolak2010}. For cold pure branches only, $\chi_{\rm all}=\chi_\infty=h_2[(1+V)/2]$ also holds with tunneling. A plateau of average fragment information is not itself proof of a successful disjoint partition or observer consensus.

\emph{Robustness of the redundancy.} Figure~\ref{fig:Sdarwinrobust} compares partitions of the cold pure-dephasing bath. Coarser partitions combine the same band weights. Both random and contiguous fragments exhibit information plateaus, but their redundancy estimates differ; a numerical redundancy is therefore meaningful only for a stated partition, sampling convention, time, and threshold. All panels use the same band-aligned integration and $4096$ permutations as Sec.~III; contiguous fragments are windows of adjacent bands.

\section{Robustness}
\label{si:robust}
\begin{figure}[!t]
\centering
\includegraphics[width=\columnwidth]{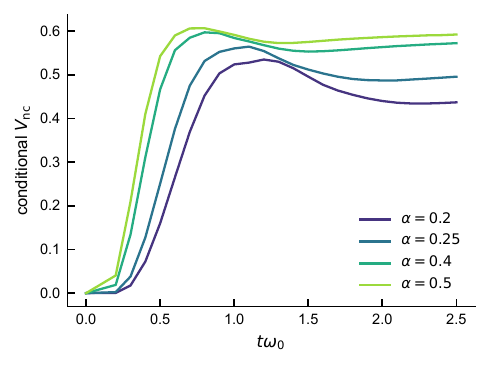}
\caption{\textbf{Coupling robustness.} Conditional $\Vnc^{(+x)}(t)$ for $\alpha=0.2,0.25,0.4$, and $0.5$ at $T=0$, $s=0.5$, and $\Delta=\omega_0$.}
\label{fig:Salpha}
\end{figure}

Figure~\ref{fig:Salpha} tests whether the conditional nonclassicality depends on a narrowly tuned coupling. At fixed $s=0.5$, $T=0$, and $\Delta=\omega_0$, conditional Wigner reconstructions for $\alpha=0.2,0.25,0.4$, and $0.5$ show a Wigner-negative cat at every tested coupling. The peak $\Vnc^{(+x)}$ rises moderately, $0.535,0.564,0.597$, and $0.606$, while its time moves from $t\omega_0\approx1.2$ at $\alpha=0.2$ to $\approx0.8$ at $\alpha=0.5$. Thus stronger coupling accelerates branch separation and modestly increases the negative volume within this range. This sweep uses the unfiltered cold mode with displacement operators obtained by exponentiating the truncated generator (Sec.~\ref{si:wigner}). At $\alpha=0.4$, a separate evolution reconstructed with projected displacements (Fig.~\ref{fig:law}) agrees with it within $9\times10^{-4}$ at the common sampled times.

\section{Convergence and error budget}
\label{si:conv}
Table~\ref{tab:Sconv} lists the energy drift, bond dimension, Fock-edge population, boundary occupation, and Wigner normalization of the tunneling trajectories. At $T/\omega_0=1$, increasing $D_{\max}$ to $600$, reducing the time step to $\Delta t\omega_0=0.005$, or raising the minimum local Fock dimension to $20$ changes $\avg{\sigma_x}$ by at most $1.4\times10^{-6}$, $5.6\times10^{-5}$, and $6.9\times10^{-7}$, respectively. These parameter tests concern the qubit coherence, while the grid refinements of Sec.~\ref{si:wigner} test the Wigner reconstruction; convergence of the detected Wigner function with respect to the evolution parameters and spectral cutoff remains to be established. All tunneling results refer to the simulated interval $0\le t\omega_0\le2.5$.

\section{Supplementary videos}
\label{si:videos}
Supplementary Video~1 shows the unfiltered $+x$-conditioned cold mode of Figs.~\ref{fig:emergence} and~\ref{fig:law}; Videos~2 and~3 show the detector band of Fig.~\ref{fig:thermal} at $T/\omega_0=0.5$ and $1$. All use $\Delta=\omega_0$, $s=0.5$, and $\alpha=0.4$. Each shows the $126$ time steps $t\omega_0=0,0.02,\ldots,2.5$ at six time steps per second, in the view $|q|,|p|\le6$ with a linear color scale $\pm2/\pi$. Peaks are $0.598$, $0.105$, and $0.0050$ at $t\omega_0=0.84$, $0.46$, and $0.28$. Video~1 ends at $0.572$; Videos~2 and~3 fall below the $10^{-4}$ reporting threshold after $t\omega_0=1.06$ and $0.44$.

\makeatletter
\let\original@label\label
\def\label#1{%
  \def\label@argument{#1}%
  \def\lastbib@label{LastBibItem}%
  \ifx\label@argument\lastbib@label
    \original@label{LastBibItemSI}%
  \else
    \original@label{#1}%
  \fi
}
\putbib[reference]
\let\label\original@label
\makeatother
\end{bibunit}

\end{document}